\documentstyle[12pt,dina4]{article}
\newcounter{saveeqn}
\newcommand{\alpheqn}[1]{
\addtocounter{equation}{1}%
\newcounter{#1}%
\setcounter{saveeqn}{\value{equation}}%
\setcounter{#1}{\value{equation}}%
\setcounter{equation}{0}%
\renewcommand{\theequation}{\mbox{\arabic{saveeqn}\alph{equation}}}}
\newcommand{\reseteqn}{\setcounter{equation}{\value{saveeqn}}%
\renewcommand{\theequation}{\arabic{equation}}}
\def\e{\mbox{\rm e}}

\def\title #1 {
   \headsep=1.0in
   \baselineskip=30pt
			\begin{center}
   {\titlebold #1}
   \end{center}
			\vskip .75in }
\def\author #1 {
   \baselineskip=30pt
   \begin{center}
   {\timeslarge #1}
   \end{center}
			\vskip .25in }
\def\address #1 {
   \baselineskip=24pt
   \begin{center}
   {\timesitalic #1}
   \end{center}
   \vskip 1.0in }

\def\Re {\mathop{\rm Re}}

\def\disp {\displaystyle}

\def\conj #1 {\overline #1}

\def\be {\begin{equation}}
\def\ee {\end{equation}}

\def\ba {\begin{array}}
\def\ea {\end{array}}
\def\bea {\begin{eqnarray}}
\def\eea {\end{eqnarray}}

\def\et {$$}

\def\etn {$$}

\def\ett {$$}

\def\ettn{$$}
\def\non  {\nonumber}
\def\eqn#1 {(\ref{#1}) }

\newdimen\twoeqncolwidth
\newdimen\twoeqncolwidtha
\newdimen\twoeqncolwidthb
\newdimen\twoeqncolsep
\newdimen\twoeqnlinset
\def\twoeqn#1&#2\et{
   \hbox to\twoeqnlinset{\hfil}
   \hbox to\twoeqncolwidth{$\disp#1$\hfil}
   \hbox to\twoeqncolsep{\hfil}
   \hbox to\twoeqncolwidth{$\disp#2$\hfil}\eqno{\rm (\theequation)}$$}
\def\twoeqnt#1&#2\ett{
   \hbox to\twoeqnlinset{\hfil}
   \hbox to\twoeqncolwidtha{$\disp#1$\hfil}
   \hbox to\twoeqncolsep{\hfil}
   \hbox to\twoeqncolwidthb{$\disp#2$\hfil}\eqno{\rm (\theequation)}$$}
\def\twoeqnn#1&#2\etn{
   \hbox to\twoeqnlinset{\hfil}
   \hbox to\twoeqncolwidth{$\disp#1$\hfil}
   \hbox to\twoeqncolsep{\hfil}
   \hbox to\twoeqncolwidth{$\disp#2$\hfil}\eqno\phantom{\rm
(\theequation)}$$}
\def\twoeqntn#1&#2\ettn{
   \hbox to\twoeqnlinset{\hfil}
   \hbox to\twoeqncolwidtha{$\disp#1$\hfil}
   \hbox to\twoeqncolsep{\hfil}
   \hbox to\twoeqncolwidthb{$\disp#2$\hfil}\eqno\phantom{\rm
(\theequation)}$$}

\def\rawpicture #1 by #2 (#3){
  \vbox to #2{
    \hrule width #1 height 0pt depth 0pt
    \vfill
    \special{picture #3}     }
  }
\def\scaledpicture #1 by #2 (#3 scaled #4){{
  \dimen0=#1 \dimen1=#2
  \divide\dimen0 by 1000 \multiply\dimen0 by #4
  \divide\dimen1 by 1000 \multiply\dimen1 by #4
  \rawpicture \dimen0 by \dimen1 (#3 scaled #4)}
  }

\def\beginparmode{\endmode
  \begingroup \def\endmode{\par\endgroup}}
\let\endmode=\par
{\obeylines\gdef\
{}}

\def\body			  {\beginparmode}
\def\head#1{			  \goodbreak\vskip 0.5truein	  {\immediate\write16{#1}
      \uppercase{#1}\par}
   \nobreak\vskip 0.25truein\nobreak}

\def\itemj{\par\hang\textindent}
\def\beginitems{
\par\medskip\bgroup\def\i##1 {\itemj{##1}}\def\ii##1 {\itemitem{##1}}
\leftskip=36pt\parskip=0pt}
\def\enditems{\par\egroup}

\def\beneathrel#1\under#2{\mathrel{\mathop{#2}\limits_{#1}}}

\def\refto#1{[#1]}
\def\references			  {\head{REFERENCES}		   \beginparmode
   \frenchspacing \parindent=0pt \leftskip=1truecm
   \parskip=8pt plus 3pt \everypar{\hangindent=\parindent}}

\gdef\refis#1{\itemj{#1.\ }}
\gdef\journal#1, #2, #3, 1#4#5#6{		    {\sl #1~}{\bf #2} (1#4#5#6), #3 }

\def\annp{\journal Ann. Phys. (N.Y.), }

\def\Annp{\journal Ann. Physik, }

\def\aspm{\journal Advanced Studies in Pure Mathematics, }

\def\cmp{\journal Comm. Math. Phys., }

\def\eurolett{\journal Europhysics Lett., }

\def\ijmpa{\journal Int. J. Mod. Phys. A, }

\def\ijmpb{\journal Int. J. Mod. Phys. B, }

\def\jappp{\journal J. Appl. Phys., }

\def\jphc{\journal J. Physique C, }

\def\jpI{\journal J. Physique I, }

\def\jpcoll{\journal J. Physique Coll, }

\def\jcr{\journal J. Chem. Res., }

\def\jetp{\journal Sov. Phys. JETP, }

\def\jetpl{\journal JETP Lett., }

\def\jpj{\journal J. Phys. Soc. Japan, }

\def\jmp{\journal J. Math. Phys., }

\def\jpa{\journal J. Phys. A, }

\def\jpc{\journal J. Phys. C, }

\def\jpcon{\journal J. Phys.: Condens. Matter, }

\def\ptp{\journal Prog. Theor. Phys., }

\def\jetp{\journal Sov. Phys. JETP, }

\def\jsp{\journal J. Stat. Phys., }

\def\lmp{\journal Lett. Math. Phys., }

\def\lnp{\journal Lecture Notes in Physics, }

\def\mpla{\journal Mod. Phys. Lett. A, }

\def\npb{\journal Nucl. Phys. B, }

\def\physica{\journal Physica, }

\def\pla{\journal Phys. Lett. A, }

\def\plb{\journal Phys. Lett. B, }

\def\prep{\journal Physics Reports, }

\def\pra{\journal Phys. Rev. A, }

\def\prb{\journal Phys. Rev. B, }

\def\prl{\journal Phys. Rev. Lett., }

\def\prs{\journal Proc. Roy. Soc. (London) A, }

\def\pr{\journal Phys. Rev., }

\def\rmp{\journal Rev. Mod. Phys., }

\def\sjnp{\journal Sov. J. Nucl. Phys., }

\def\tmp{\journal Theor. Math. Phys., }

\def\zpb{\journal Z. Phys. B, }

\def\zp{\journal Z. Phys., }

\def\reff#1{Ref.~#1}			\def\Reff#1{Ref.~#1}			\def\[#1]{[\refcite{#1}]}
\def\refcite#1{{#1}}

				\def\(#1){(\call{#1})}
\def\call#1{{#1}}
\def\taghead#1{}
\def\frac#1#2{{#1 \over #2}}

\def\sla{\raise.15ex\hbox{$/$}\kern-.57em}
\def\leaderfill{\leaders\hbox to 1em{\hss.\hss}\hfill}
\def\twiddle{\lower.9ex\rlap{$\kern-.1em\scriptstyle\sim$}}
\def\bigtwiddle{\lower1.ex\rlap{$\sim$}}
\def\gtwid{\mathrel{\raise.3ex\hbox{$>$\kern-.75em\lower1ex\hbox{$\sim$}}}}
\def\ltwid{\mathrel{\raise.3ex\hbox{$<$\kern-.75em\lower1ex\hbox{$\sim$}}}}
\def\square{\kern1pt\vbox{\hrule height 1.2pt\hbox{\vrule width 1.2pt\hskip
3pt
   \vbox{\vskip 6pt}\hskip 3pt\vrule width 0.6pt}\hrule height
0.6pt}\kern1pt}
\def\tdot#1{\mathord{\mathop{#1}\limits^{\kern2pt\ldots}}}

\def\pmb#1{\setbox0=\hbox{#1}  \kern-.025em\copy0\kern-\wd0
  \kern  .05em\copy0\kern-\wd0
  \kern-.025em\raise.0433em\box0 }

\def\refto#1{[#1]}
\def\references			  {\section*{References}		   \beginparmode
   \frenchspacing \parindent=0pt \leftskip=1truecm
   \parskip=8pt plus 3pt \everypar{\hangindent=\parindent}}

\def\endreferences{\body}

\def\reff#1{Ref.~#1}			\def\Reff#1{Ref.~#1}			\def\[#1]{[\refcite{#1}]}
\def\refcite#1{{#1}}
\def\refeq#1{(\ref{#1})}

\catcode`@=11
\newcount\r@fcount \r@fcount=0
\newcount\r@fcurr
\immediate\newwrite\reffile
\newif\ifr@ffile\r@ffilefalse
\def\w@rnwrite#1{\ifr@ffile\immediate\write\reffile{#1}\fi\message{#1}}

\def\writer@f#1>>{}
\def\referencefile{  \r@ffiletrue\immediate\openout\reffile=\jobname.ref
\def\writer@f##1>>{\ifr@ffile\immediate\write\reffile    {\noexpand\refis{##1}
= \csname r@fnum##1\endcsname =
\expandafter\expandafter\expandafter\strip@t\expandafter     \meaning\csname
r@ftext\csname r@fnum##1\endcsname\endcsname}\fi}  \def\strip@t##1>>{}}

\def\citeall#1{\xdef#1##1{#1{\noexpand\refcite{##1}}}}
\def\refcite#1{\each@rg\citer@nge{#1}}

\def\each@rg#1#2{{\let\thecsname=#1\expandafter\first@rg#2,\end,}}
\def\first@rg#1,{\thecsname{#1}\apply@rg}
\def\apply@rg#1,{\ifx\end#1\let\next=\relax\else,\thecsname{#1}\let\next=\apply@rg\fi\next}
\def\citer@nge#1{\citedor@nge#1-\end-}
\def\citer@ngeat#1\end-{#1}
\def\citedor@nge#1-#2-{\ifx\end#2\r@featspace#1
\else\citel@@p{#1}{#2}\citer@ngeat\fi}
\def\citel@@p#1#2{\ifnum#1>#2{\errmessage{Reference range #1-#2\space is
bad.}
    \errhelp{If you cite a series of references by the notation M-N, then M
and
    N must be integers, and N must be greater than or equal to M.}}\else
{\count0=#1\count1=#2\advance\count1
by1\relax\expandafter\r@fcite\the\count0,  \loop\advance\count0 by1\relax
\ifnum\count0<\count1,\expandafter\r@fcite\the\count0,  \repeat}\fi}

\def\r@featspace#1#2 {\r@fcite#1#2,}	\def\r@fcite#1,{\ifuncit@d{#1}
\newr@f{#1}    \expandafter\gdef\csname r@ftext\number\r@fcount\endcsname
              {\message{Reference #1 to be supplied.}
\writer@f#1>>#1 to be supplied.\par} \fi \csname r@fnum#1\endcsname}
\def\ifuncit@d#1{\expandafter\ifx\csname
r@fnum#1\endcsname\relax}\def\newr@f#1{\global\advance\r@fcount by1
\expandafter\xdef\csname r@fnum#1\endcsname{\number\r@fcount}}

\let\r@fis=\refis			\def\refis#1#2#3\par{\ifuncit@d{#1}   \newr@f{#1}
\w@rnwrite{Reference #1=\number\r@fcount\space is not cited up to
now.}\fi  \expandafter\gdef\csname r@ftext\csname r@fnum#1\endcsname\endcsname
{\writer@f#1>>#2#3\par}}

\def\ignoreuncited{   \def\refis##1##2##3\par{\ifuncit@d{##1}
\else\expandafter\gdef\csname r@ftext\csname
r@fnum##1\endcsname\endcsname     {\writer@f##1>>##2##3\par}\fi}}

\def\r@ferr{\endreferences\errmessage{I was expecting to see
\noexpand\endreferences before now;  I have inserted it here.}}
\let\r@ferences=\references
\def\references{\r@ferences\def\endmode{\r@ferr\par\endgroup}}

\let\endr@ferences=\endreferences
\def\endreferences{\r@fcurr=0  {\loop\ifnum\r@fcurr<\r@fcount
\advance\r@fcurr by
1\relax\expandafter\r@fis\expandafter{\number\r@fcurr}    \csname
r@ftext\number\r@fcurr\endcsname  \repeat}\gdef\r@ferr{}\endr@ferences}

\let\r@fend=\endpaper\gdef\endpaper{\ifr@ffile
\immediate\write16{Cross References written on []\jobname.REF.}\fi\r@fend}

\catcode`@=12

\citeall\refto		\citeall\reff		\citeall\Reff

\ignoreuncited

\renewcommand{\title}[1]{\large\bf \mbox{}\\ \mbox{}\\ \mbox{}\\ \mbox{}\\
     #1\bigskip\medskip\\}
\renewcommand{\author}[1]{\large #1\\ \smallskip}
\renewcommand{\address}[1]{{\narrower\normalsize\it #1\\}\bigskip}
\renewenvironment{abstract}{\narrower\small}{\par\normalsize\bigskip}

\catcode`\@=11

\font\twelvemsx=msxm10 scaled \magstep1
\font\tenmsx=msxm10
\font\sevenmsx=msxm7

\font\twelvemsy=msym10 scaled \magstep1
\font\tenmsy=msym10
\font\sevenmsy=msym7

\newfam\msxfam
\newfam\msyfam
\textfont\msxfam=\twelvemsx
\scriptfont\msxfam=\tenmsx
  \scriptscriptfont\msxfam=\sevenmsx
\textfont\msyfam=\twelvemsy
\scriptfont\msyfam=\tenmsy
  \scriptscriptfont\msyfam=\sevenmsy

\def\hexnumber@#1{\ifcase#1 0\or1\or2\or3\or4\or5\or6\or7\or8\or9\or
	A\or B\or C\or D\or E\or F\fi }

\font\twelveeuf=eufm10 scaled \magstep1
\font\teneuf=eufm10
\font\seveneuf=eufm7

\newfam\euffam
\textfont\euffam=\twelveeuf
\scriptfont\euffam=\teneuf
\scriptscriptfont\euffam=\seveneuf
\def\frak{\relaxnext@\ifmmode\let\next\frak@\else
 \def\next{\Err@{Use \string\frak\space only in math mode}}\fi\next}
\def\goth{\relaxnext@\ifmmode\let\next\frak@\else
 \def\next{\Err@{Use \string\goth\space only in math mode}}\fi\next}
\def\frak@#1{{\frak@@{#1}}}
\def\frak@@#1{\noaccents@\fam\euffam#1}

\edef\msx@{\hexnumber@\msxfam}
\edef\msy@{\hexnumber@\msyfam}

\mathchardef\boxdot="2\msx@00
\mathchardef\boxplus="2\msx@01
\mathchardef\boxtimes="2\msx@02
\mathchardef\square="0\msx@03
\mathchardef\blacksquare="0\msx@04
\mathchardef\centerdot="2\msx@05
\mathchardef\lozenge="0\msx@06
\mathchardef\blacklozenge="0\msx@07
\mathchardef\circlearrowright="3\msx@08
\mathchardef\circlearrowleft="3\msx@09
\mathchardef\rightleftharpoons="3\msx@0A
\mathchardef\leftrightharpoons="3\msx@0B
\mathchardef\boxminus="2\msx@0C
\mathchardef\Vdash="3\msx@0D
\mathchardef\Vvdash="3\msx@0E
\mathchardef\vDash="3\msx@0F
\mathchardef\twoheadrightarrow="3\msx@10
\mathchardef\twoheadleftarrow="3\msx@11
\mathchardef\leftleftarrows="3\msx@12
\mathchardef\rightrightarrows="3\msx@13
\mathchardef\upuparrows="3\msx@14
\mathchardef\downdownarrows="3\msx@15
\mathchardef\upharpoonright="3\msx@16

\mathchardef\downharpoonright="3\msx@17
\mathchardef\upharpoonleft="3\msx@18
\mathchardef\downharpoonleft="3\msx@19
\mathchardef\rightarrowtail="3\msx@1A
\mathchardef\leftarrowtail="3\msx@1B
\mathchardef\leftrightarrows="3\msx@1C
\mathchardef\rightleftarrows="3\msx@1D
\mathchardef\Lsh="3\msx@1E
\mathchardef\Rsh="3\msx@1F
\mathchardef\rightsquigarrow="3\msx@20
\mathchardef\leftrightsquigarrow="3\msx@21
\mathchardef\looparrowleft="3\msx@22
\mathchardef\looparrowright="3\msx@23
\mathchardef\circeq="3\msx@24
\mathchardef\succsim="3\msx@25
\mathchardef\gtrsim="3\msx@26
\mathchardef\gtrapprox="3\msx@27
\mathchardef\multimap="3\msx@28
\mathchardef\therefore="3\msx@29
\mathchardef\because="3\msx@2A
\mathchardef\doteqdot="3\msx@2B

\mathchardef\triangleq="3\msx@2C
\mathchardef\precsim="3\msx@2D
\mathchardef\lesssim="3\msx@2E
\mathchardef\lessapprox="3\msx@2F
\mathchardef\eqslantless="3\msx@30
\mathchardef\eqslantgtr="3\msx@31
\mathchardef\curlyeqprec="3\msx@32
\mathchardef\curlyeqsucc="3\msx@33
\mathchardef\preccurlyeq="3\msx@34
\mathchardef\leqq="3\msx@35
\mathchardef\leqslant="3\msx@36
\mathchardef\lessgtr="3\msx@37
\mathchardef\backprime="0\msx@38
\mathchardef\risingdotseq="3\msx@3A
\mathchardef\fallingdotseq="3\msx@3B
\mathchardef\succcurlyeq="3\msx@3C
\mathchardef\geqq="3\msx@3D
\mathchardef\geqslant="3\msx@3E
\mathchardef\gtrless="3\msx@3F
\mathchardef\sqsubset="3\msx@40
\mathchardef\sqsupset="3\msx@41
\mathchardef\vartriangleright="3\msx@42
\mathchardef\vartriangleleft="3\msx@43
\mathchardef\trianglerighteq="3\msx@44
\mathchardef\trianglelefteq="3\msx@45
\mathchardef\bigstar="0\msx@46
\mathchardef\between="3\msx@47
\mathchardef\blacktriangledown="0\msx@48
\mathchardef\blacktriangleright="3\msx@49
\mathchardef\blacktriangleleft="3\msx@4A
\mathchardef\vartriangle="0\msx@4D
\mathchardef\blacktriangle="0\msx@4E
\mathchardef\triangledown="0\msx@4F
\mathchardef\eqcirc="3\msx@50
\mathchardef\lesseqgtr="3\msx@51
\mathchardef\gtreqless="3\msx@52
\mathchardef\lesseqqgtr="3\msx@53
\mathchardef\gtreqqless="3\msx@54
\mathchardef\Rrightarrow="3\msx@56
\mathchardef\Lleftarrow="3\msx@57
\mathchardef\veebar="2\msx@59
\mathchardef\barwedge="2\msx@5A
\mathchardef\doublebarwedge="2\msx@5B
\mathchardef\angle="0\msx@5C
\mathchardef\measuredangle="0\msx@5D
\mathchardef\sphericalangle="0\msx@5E
\mathchardef\varpropto="3\msx@5F
\mathchardef\smallsmile="3\msx@60
\mathchardef\smallfrown="3\msx@61
\mathchardef\Subset="3\msx@62
\mathchardef\Supset="3\msx@63
\mathchardef\Cup="2\msx@64

\mathchardef\Cap="2\msx@65

\mathchardef\curlywedge="2\msx@66
\mathchardef\curlyvee="2\msx@67
\mathchardef\leftthreetimes="2\msx@68
\mathchardef\rightthreetimes="2\msx@69
\mathchardef\subseteqq="3\msx@6A
\mathchardef\supseteqq="3\msx@6B
\mathchardef\bumpeq="3\msx@6C
\mathchardef\Bumpeq="3\msx@6D
\mathchardef\lll="3\msx@6E

\mathchardef\ggg="3\msx@6F

\mathchardef\circledS="0\msx@73
\mathchardef\pitchfork="3\msx@74
\mathchardef\dotplus="2\msx@75
\mathchardef\backsim="3\msx@76
\mathchardef\backsimeq="3\msx@77
\mathchardef\complement="0\msx@7B
\mathchardef\intercal="2\msx@7C
\mathchardef\circledcirc="2\msx@7D
\mathchardef\circledast="2\msx@7E
\mathchardef\circleddash="2\msx@7F
\def\ulcorner{\delimiter"4\msx@70\msx@70 }
\def\urcorner{\delimiter"5\msx@71\msx@71 }
\def\llcorner{\delimiter"4\msx@78\msx@78 }
\def\lrcorner{\delimiter"5\msx@79\msx@79 }
\def\yen{\mathhexbox\msx@55 }
\def\checkmark{\mathhexbox\msx@58 }
\def\circledR{\mathhexbox\msx@72 }
\def\maltese{\mathhexbox\msx@7A }
\mathchardef\lvertneqq="3\msy@00
\mathchardef\gvertneqq="3\msy@01
\mathchardef\nleq="3\msy@02
\mathchardef\ngeq="3\msy@03
\mathchardef\nless="3\msy@04
\mathchardef\ngtr="3\msy@05
\mathchardef\nprec="3\msy@06
\mathchardef\nsucc="3\msy@07
\mathchardef\lneqq="3\msy@08
\mathchardef\gneqq="3\msy@09
\mathchardef\nleqslant="3\msy@0A
\mathchardef\ngeqslant="3\msy@0B
\mathchardef\lneq="3\msy@0C
\mathchardef\gneq="3\msy@0D
\mathchardef\npreceq="3\msy@0E
\mathchardef\nsucceq="3\msy@0F
\mathchardef\precnsim="3\msy@10
\mathchardef\succnsim="3\msy@11
\mathchardef\lnsim="3\msy@12
\mathchardef\gnsim="3\msy@13
\mathchardef\nleqq="3\msy@14
\mathchardef\ngeqq="3\msy@15
\mathchardef\precneqq="3\msy@16
\mathchardef\succneqq="3\msy@17
\mathchardef\precnapprox="3\msy@18
\mathchardef\succnapprox="3\msy@19
\mathchardef\lnapprox="3\msy@1A
\mathchardef\gnapprox="3\msy@1B
\mathchardef\nsim="3\msy@1C
\mathchardef\ncong="3\msy@1D

\mathchardef\varsubsetneq="3\msy@20
\mathchardef\varsupsetneq="3\msy@21
\mathchardef\nsubseteqq="3\msy@22
\mathchardef\nsupseteqq="3\msy@23
\mathchardef\subsetneqq="3\msy@24
\mathchardef\supsetneqq="3\msy@25
\mathchardef\varsubsetneqq="3\msy@26
\mathchardef\varsupsetneqq="3\msy@27
\mathchardef\subsetneq="3\msy@28
\mathchardef\supsetneq="3\msy@29
\mathchardef\nsubseteq="3\msy@2A
\mathchardef\nsupseteq="3\msy@2B
\mathchardef\nparallel="3\msy@2C
\mathchardef\nmid="3\msy@2D
\mathchardef\nshortmid="3\msy@2E
\mathchardef\nshortparallel="3\msy@2F
\mathchardef\nvdash="3\msy@30
\mathchardef\nVdash="3\msy@31
\mathchardef\nvDash="3\msy@32
\mathchardef\nVDash="3\msy@33
\mathchardef\ntrianglerighteq="3\msy@34
\mathchardef\ntrianglelefteq="3\msy@35
\mathchardef\ntriangleleft="3\msy@36
\mathchardef\ntriangleright="3\msy@37
\mathchardef\nleftarrow="3\msy@38
\mathchardef\nrightarrow="3\msy@39
\mathchardef\nLeftarrow="3\msy@3A
\mathchardef\nRightarrow="3\msy@3B
\mathchardef\nLeftrightarrow="3\msy@3C
\mathchardef\nleftrightarrow="3\msy@3D
\mathchardef\divideontimes="2\msy@3E
\mathchardef\varnothing="0\msy@3F
\mathchardef\nexists="0\msy@40
\mathchardef\mho="0\msy@66
\mathchardef\eth="0\msy@67
\mathchardef\eqsim="3\msy@68
\mathchardef\beth="0\msy@69
\mathchardef\gimel="0\msy@6A
\mathchardef\daleth="0\msy@6B
\mathchardef\lessdot="3\msy@6C
\mathchardef\gtrdot="3\msy@6D
\mathchardef\ltimes="2\msy@6E
\mathchardef\rtimes="2\msy@6F
\mathchardef\shortmid="3\msy@70
\mathchardef\shortparallel="3\msy@71
\mathchardef\smallsetminus="2\msy@72
\mathchardef\thicksim="3\msy@73
\mathchardef\thickapprox="3\msy@74
\mathchardef\approxeq="3\msy@75
\mathchardef\succapprox="3\msy@76
\mathchardef\precapprox="3\msy@77
\mathchardef\curvearrowleft="3\msy@78
\mathchardef\curvearrowright="3\msy@79
\mathchardef\digamma="0\msy@7A
\mathchardef\varkappa="0\msy@7B
\mathchardef\hslash="0\msy@7D
\mathchardef\hbar="0\msy@7E
\mathchardef\backepsilon="3\msy@7F
\def\Bbb@@#1{\fam\msyfam#1}
\def\Bbb@#1{{\Bbb@@{#1}}}
\def\Bbb{\Bbb@}

\catcode`\@=12

\font\twelvemsx=msxm10 scaled \magstep1
\font\twelvemsy=msym10 scaled \magstep1
\font\twelveeuf=eufm10 scaled \magstep1
\textfont\msxfam=\twelvemsx
\textfont\msyfam=\twelvemsy
\textfont\euffam=\twelveeuf
\def\frak#1{\mbox{\twelveeuf #1}}

\def\phi{\varphi}
\def\-{{\bf --}}

\newcommand{\cc}{{\hbox to 8pt {\hfill\vrule height 6.5pt \kern-2.6pt {\rm C}
                  \hfill}}}

\newcounter{num}

\def\e{{\rm e}}

\def\qLam#1#2{q^{#1}\left({i\over 2}#2 \eta\right)}
\def\qLam2#1#2{q^{#1}\left({i\over 2}#2 {\eta\over 2}\right)}

\begin{document}
\begin{center}

\titlepage

\title{Exact solution of new integrable nineteen-vertex models and
quantum spin-1 chains\footnote{
Work performed within the research program of the
Sonderforschungsbereich 341, K\"oln-Aachen-J\"ulich}}

\vskip1cm

\author{A.~Kl\"umper, S.~I.~Matveenko\footnote{Permanent address:
L.D.Landau Institute for Theoretical Physics, Kosygina str. 2,
GSP-1, 117940, Moscow, Russia; Email: matveen@cpd.landau.free.net},
J.~Zittartz}

\address{Institut f\"ur Theoretische Physik,
Universit\"at zu K\"oln, Z\"ulpicher Str. 77,\\
D-50937 K\"oln, Germany.
\footnote{Email: kluemper@thp.uni-koeln.de, zitt@thp.uni-koeln.de}}
\end{center}

\vskip1cm

\begin{abstract}
New exactly solvable nineteen vertex models and related quantum spin-1
chains  are solved. Partition functions, excitation energies,
 correlation lengths, and critical exponents are calculated. It is argued
that one of the non-critical Hamiltonians is a realization of an integrable
Haldane system. The finite-size spectra of the critical Hamiltonians
deviate in their structure from standard predictions by conformal invariance.
\end{abstract}


\today

\bibliographystyle{alpha}
\section{Introduction}

\indent

In a recent paper \refto{Idz94} a complete list of exactly solvable cases
was presented for the
three-state vertex model with ice rule and certain symmetries.
As a criterion for integrability the ``additive" Yang-Baxter equation was used
 and solved exhaustively for the Boltzmann weights in the considered class of
models.
The list in \refto{Idz94} comprises several well-known models, but also four
non-trivial new ones. The solution to these models was not given in
\refto{Idz94}.
It is the purpose of the present paper to present the exact solution of
the four models, namely the calculation of the
thermodynamic properties, i.e. partition functions, correlation lengths,
critical exponents, etc. At the same time we solve the associated quantum
spin-1 chains.

The vertex models considered in this paper are defined on a square lattice
where spin variables are placed on the bonds. Each spin may take three values,
say 0 or $\pm 1$, and there are interaction energies associated with each
vertex
 depending on the local spin configuration. The corresponding local Boltzmann
weight is denoted by $R^{\mu\,\alpha}_{\nu\,\beta}$ for a spin configuration
$\mu, \nu, \alpha,$ and $\beta$ on the left, right, lower, and upper bond of
the
vertex, respectively.

The partition function for a lattice of size $N \times L$ with periodic
boundary conditions is given by
\begin{equation}
Z = {\rm Tr} \;T^L,
\end{equation}
where the transfer-matrix  $T$ is the product of the Boltzmann
weights in a row:
\begin{equation}
T^{\alpha_1, \ldots \alpha_N}_{\beta_1 , \ldots \beta_N} =
\sum_{\mu_1 , \ldots \mu_N} R^{\mu_1\, \alpha_1}_{\mu_2 \, \beta_1}
 R^{\mu_2\, \alpha_2}_{\mu_3 \, \beta_2}\ldots
 R^{\mu_N\, \alpha_N}_{\mu_1 \, \beta_N}.\label{transfer}
\end{equation}

As in \refto{Idz94} we impose the ice-rule
\begin{equation}
\alpha + \mu = \beta + \nu
\label{m1}
\end{equation}
which leads to nineteen allowed, i.e. non-zero, Boltzmann weights,
while the further symmetries
\begin{equation}
R^{\mu \alpha}_{\nu \beta} = R^{-\mu -\alpha}_{-\nu -\beta} =
R^{ \alpha \mu}_{ \beta \nu} = R^{\nu \beta}_{\mu \alpha}.
\label{m2}
\end{equation}
reduce them to only 7 independent weights:
\begin{eqnarray}
a = R^{1\,1}_{1\,1},&b = R^{-1\,1}_{-1\,1},&c = R^{-1\,1}_{1\,-1} \nonumber \\
e = R^{1\,0}_{0\,1},&g = R^{1\,-1}_{0\,0}, &p = R^{0\,1}_{0\,1}, \nonumber \\
d = R^{0\,0}_{0\,0}.&                      &  \label{m3}
\end{eqnarray}

In the following we shall consider the four new models which in \refto{Idz94}
have been derived as the solutions $ \# 8, \# 9,
\# 2$, and $\# 3$  of the Yang-Baxter equation and which we label as
I to IV.
The four models are defined by their respective (not normalized)
Boltzmann weights which all depend on the spectral parameter $u$.
Models I and II have no further variable interaction parameter and
are parametrized (with slight changes as compared to \refto{Idz94})
as follows. With $U  = \exp {u}$ we have the weights:

\medskip

{Model I}:
$$
a = {4-U^4\over 3U^2},\quad b={2(U^4 -1)\over 3U^2},\quad
c = {U^4 + 2\over 3U^2},
$$
\begin{equation}
e = \frac{a}{U},\quad g = \frac{U}{\sqrt{2}} b, \quad p = 0,\quad d = 1.
\label{m5}
\end{equation}

{Model II}:
$$
a = \frac{\alpha (\alpha^2 - U^4)}{U^2(U^4 + \alpha)},\quad b = \alpha
U^2 \frac{U^4 - 1}{U^4 + \alpha}, \quad c = \frac{\alpha^2 U^2}{U^4 +
\alpha}
$$
\begin{equation}
e = Ua,\quad g = \frac{\sqrt{\alpha}}{U} b,\quad p = 0,\quad d = 1,
\label{m6}
\end{equation}
where $\alpha = \frac{\sqrt{5} + 1}{2}$.

In the physical region all Boltzmann weights must be positive which leads to
\be
0\le {\rm Re }\, u \le \lambda, \qquad
\lambda = \Biggl\{ \begin{array}{ll}
                   \ln \sqrt{2},     & \mbox{for model I},\\
                   \ln \sqrt \alpha, & \mbox{for  model II.}
                        \end{array}
               \label{physReg}
\ee

The models III and IV depend on an additional interaction parameter
besides the spectral parameter $u$ and are parametrized differently in
different regions.
\medskip

{Model III}:
We have $b=g=0$ and $a=c=d$.
With
\be
\Delta={a^2+p^2-e^2\over 2ap},\qquad (-\infty<\Delta<\infty)\label{Del}
\ee
the parametrization depends on the value of $\Delta$.

\alpheqn{modiii}

\begin{itemize}
\item $\Delta>1$:
\begin{equation}
a={\sinh(\lambda+u)\over \sinh\lambda}, \quad
p={\sinh u\over \sinh\lambda},\quad e=1,\quad \Delta=\cosh\lambda,
\label{m8a}
\end{equation}

\item $\Delta<-1$:
\begin{equation}
a={\sinh(\lambda-u)\over \sinh\lambda}, \quad
p={\sinh u\over \sinh\lambda},\quad e=1,\quad \Delta=-\cosh\lambda,
\label{m8b}
\end{equation}

\item $-1<\Delta<1$:
\begin{equation}
a={\sin(\lambda-u)\over \sin\lambda}, \quad
p={\sin u\over \sin\lambda},\quad e=1,\quad \Delta=-\cos\lambda,
\label{m8c}
\end{equation}

\end{itemize}
\reseteqn
with $\lambda\ge 0$.
\medskip

{Model IV}:
Again we have $b=g=0$ and $a=c$. However, now we require
\be
ad+p^2-e^2=0,\label{freeFer}
\ee
which will prove to be a free fermion condition. Otherwise we shall use
two different parametrizations

\alpheqn{modiv}

\begin{itemize}
\item a)
$$
a=\cosh u \pm \cosh \lambda \sinh u,\quad d=\cosh u \mp \cosh \lambda \sinh u,
$$
\be
p = \sinh \lambda \sinh u,\quad e = 1, \label{m9a}
\ee

\item b)
$$
a=\cos u \pm \sinh \lambda \sin u,\quad d=\cos u \mp \sinh \lambda \sin u,
$$
\be
p = \cosh \lambda \sin u,\quad e = 1, \label{m9b}
\ee
\end{itemize}
\reseteqn
where $\lambda\ge 0$.

This completes the parametrization of the four models.
We emphasize that all parametrizations
satisfy the ``standard initial condition"
\be
R^{\mu\, \alpha}_{\nu\, \beta}(u=0) =  \delta_{\mu\, \beta}
\delta_{\nu\, \alpha}.  \label{m10}
\ee

One of the consequences of the
Yang-Baxter equation is the existence of a family of commuting transfer
matrices $T(u)$  generated by the spectral parameter $u$
$$
[T(u), T(v)] = 0.
$$
As a result of this commutativity the eigenvalue functions $\Lambda (u)$ of
$T(u)$ possess the same analytic properties as the Boltzmann weights (i.e.
analyticity up to poles imposed by the parametrization of the weights).
It is well-known that a one-dimensional quantum spin-1 chain is
associated with each 3-state two-dimensional vertex model. Its Hamiltonian
$H$ and momentum operator $P$ are defined by
\be
\tau H = -\frac{d}
{du} \ln T \vert_{u=0},\quad P = -i\ln T(0), \label{m12}
\ee
where $\tau$ is an arbitrary positive scale factor. It turns out that
because of the standard initial condition (\ref{m10}) the Hamiltonians
are sums of local terms which are given explicitly in the following sections.

The plan of this paper is as follows. In section 2 we solve the models I and
II.
We calculate the partition function and the correlation
lengths by employing analyticity of the eigenvalues
$\Lambda (u)$ and an important inversion identity. The models will
turn out to be non-critical, independently of the spectral parameter $u$
as all $T(u)$ commute for different $u$. In subsection 2.3 we derive the
associated quantum spin-1 Hamiltonian and determine its spectrum from
the spectrum of $T(u)$.

In Section 3 the models III and IV are analyzed. They are mapped to the
symmetric six-vertex model and a free fermion model
where the interaction parameter $\lambda$ plays the role of
a crossing  parameter or the chemical potential, respectively.
Depending on  $\lambda$ the models show different behaviour: there are
non-critical regimes with ferro- and antiferro-magnetic order, and a
critical antiferro-magnetic regime.\footnote[1]{In the usual terminology
of the six-vertex model one uses ``ferro- (antiferro-) electric
order" rather than ``ferro- (antiferro-) magnetic order".} We also determine
the spectrum of the associated one-dimensional Hamiltonians.

Section 4 contains a discussion.

\section{Solution of models I and II}

\indent

To solve models I and II we shall apply the analytic method developed
in \refto{KlumBiq} which uses functional
equations and avoids the more cumbersome Bethe ansatz.
We also refer to \refto{KlumBiq} for more detailed proofs and
derivations of corresponding results.
A well known
consequence of the Yang-Baxter equation for $R$-matrices satisfying the
standard initial condition (\ref{m10}) is the unitarity property
\be
\sum_{\gamma, \delta} R_{\delta\gamma}^{\mu \alpha} (u)
R^{\gamma\delta}_{\nu\beta} (-u) =
\phi (u) \delta_{\alpha \nu} \delta_{\mu \beta}, \label{m13}
\ee
where
\be
\phi (u) = a(u) a(-u).\label{m14}
\ee
Explicitly we obtain
\begin{eqnarray}
\mbox{model I:\ }\quad  \phi(u)&=& \frac{1}{9}(4-U^4)(4- \frac{1}{U^4})
                                           \nonumber \\
\mbox{model II:}\quad \phi(u)&=& -\alpha^3 \frac{(U^4-\alpha^2)
                                     (U^4-\alpha^{-2})}
                              {(U^4+\alpha)(U^4+\alpha^{-1})}. \label{m15}
\end{eqnarray}

Furthermore we have the additional crossing symmetry
\be
R_{\nu\;\beta}^{\mu\;\alpha} (u) = R_{-\alpha\;\nu}^{-\beta\;\mu}
(\lambda - u) \label{m16}
\ee
where the crossing parameter $\lambda$ is given in \refeq{physReg}.

As shown in \refto{KlumBiq} the local relations (\ref{m13}) and
(\ref{m16}) imply the global inversion
relation for the transfer matrices
\be
T(u) T(u + \lambda) = {\phi (u)}^N \left[I_N + O(e^{-N})\right], \label{m18}
\ee
where $I_N$ is the identity matrix and $O(\exp{(-N)})$ is a
correction which is exponentially small
in the thermodynamic limit ($N\to\infty$).
(The correction term is identically zero for $u=0$  since the
transfer matrices $T(0)$ and $T(\lambda )$ reduce to right and left
shift operators.)

 From the local crossing symmetry (\ref{m16}) and (\ref{m2}) we also obtain
\be
T^{+}(u) = T(\lambda - u^{\ast}). \label{m19}
\ee
Relations (\ref{m18}) and (\ref{m19}) directly imply
functional equations for the eigenvalues $\Lambda (u)$
\be
\Lambda (u) \Lambda (u + \lambda) = \phi (u)^N\left[1 + O(e^{-N})\right],
\label{m20}
\ee
$$\Lambda^{\ast} (u) = \Lambda (\lambda - u^{\ast}),$$
which will be solved in the next subsections for the largest
 and next-largest eigenvalues subject to some obvious analytical properties
(and $N$ even). We note that models III and IV also satisfy \refeq{m13},
but relation \refeq{m16} does not hold. Therefore the method of solution
for these models is different (see section 3).

\subsection{Largest eigenvalue and partition function}

\indent

Here the eigenvalue
$\Lambda_{0} (u) $ which is the largest in the physical region
is determined. It is convenient to define
\be
\Psi (u) = \lim_{N \rightarrow \infty} \Lambda_{0}^{1/N} (u) \label{m21}
\ee
for which we have \\
\begin{itemize}
\item (i) analyticity in the physical region $0\leq \Re u \leq \lambda$,
no zeros therein,
\item (ii) periodicity or antiperiodicity under $u \rightarrow u +
\frac{1}{2}\pi i$,
\item (iii) inversion identity:
\be
\Psi (u) \Psi (u + \lambda) = \phi (u).\label{m22}
\ee
\end{itemize}

\noindent
For deriving the unique solution we closely follow \refto{KlumBiq}.

\medskip

{Model I}: The ansatz
\be
\Psi (u) = \frac{4}{3} F(U) F\left(\frac{\sqrt{2}}{U}\right) \label{m23}
\ee
satisfies the crossing symmetry $\Psi(u) = \Psi (\lambda - u)$
for real $\Psi$. Inserting (\ref{m23}) into (\ref{m22}) yields a
functional equation which is satisfied if only
\be
F(U) F(\sqrt{2} U) = 1 - \frac{1}{4U^4}.  \label{m24}
\ee
Solving this equation we obtain
\be
F(U) = \prod_{n=0}^{\infty} \frac{1 -\frac{1}{U^4 4^{2n+1}}}
{1 - \frac{1}{U^4 4^{2n+2}}}. \label{m25}
\ee
 The function defined by  (\ref{m23}) and (\ref{m25}) satisfies (i)
 and (ii) (it is indeed periodic). Therefore it is identical to the partition
function per site (\ref{m21}). As explained in \refto{KlumBiq} the solution is
unique.
\medskip

{Model II}: The ansatz
\be
\Psi (u) = {\alpha}^3 \frac{U^2}{\alpha + U^4} F(U) F\left(\frac{\sqrt{\alpha}}
{U}\right), \label{m26}
\ee
leads to
\be
F(U) F(\sqrt{\alpha} U) = 1 - \frac{1}{{\alpha}^2 U^4},  \label{m27}
\ee
which is solved by
\be
F(U) = \prod_{n=0}^{\infty} \frac{1 -\frac{1}{U^4 {\alpha}^{4n+2}}}
{1 - \frac{1}{U^4 {\alpha}^{4n+4}}}. \label{m28}
\ee
Again the function defined by (\ref{m26}) and (\ref{m28}) satisfies
(i) and is periodic in the sense of (ii).

These are the final results for the partition function per site. In Section
2.3 the ground state energy of the related quantum spin chain is
calculated from $\Psi (u)$.

\subsection{Next-largest eigenvalues and correlation length}

\indent

We now consider all eigenvalues of the transfer matrix for which
\be
l(u) := \lim_{N \rightarrow \infty} \frac{\Lambda (u)}{{\Lambda}_{0} (u)}
 \label{m29}
\ee
 is finite. The properties of $l(u)$ are:
\begin{itemize}
\item (i) analyticity in the physical region $0 \leq \Re\, u \leq \lambda$,
zeros are
allowed,
\item (ii) periodicity (antiperiodicity) under $u \rightarrow u +
\frac{1}{2}\pi i$
for even (odd) magnetization of the considered state,
\item (iii) inversion relation
\be l(u) l(u + \lambda) = 1, \label{m30} \ee
\item (iv) crossing symmetry \qquad $l^{\ast} (u) = l (\lambda - u^{\ast})$.
\end{itemize}

The properties (iii) and (iv) follow from \refeq{m20}. The magnetization
for a state $(\alpha_1, ..., \alpha_N)$ of vertical spins
$(\alpha_i=0,\,\pm 1)$ is defined as $M=\sum_{i=1}^N\alpha_i$ and is an
even or odd integer. This quantity is conserved by the transfer matrix $T$.
Property (ii) follows by inspecting the weights \refeq{m5}, \refeq{m6} in the
product row of $T$ \refeq{transfer}.

Applying (iii) twice we obtain $l(u + 2\lambda) = l(u)$.
Therefore, $l(u)$ has two periods $2\lambda$ and $\frac{1}{2}\pi i$.
A doubly periodic, meromorphic function is an elliptic function and
is determined by the location of its zeros $\Theta_j$ and poles
$\Theta_j+\lambda$. We obtain
\be
l(u) = \prod_{j=1}^{\nu} \sqrt{k}\, {\rm snh} \left[\frac{4K}{\pi} (u -
{\Theta}_j )\right],
 \label{m31}
\ee
where snh is the elliptic function of modulus
$k \in (0, 1)$ (see for instance \refto{Baxt82b,KlumZ88App})
which is defined by requiring that the corresponding
periods $ K(k), K^{\prime} (k)$ satisfy
\be
\frac{K^{\prime}}{K} = \frac{4\lambda}{\pi}. \label{m33}
\ee
As snh(...) is antiperiodic under $u\to u+{1\over 2}\pi i$, it follows from
(ii) that $\nu$ is even (odd) for even (odd)
magnetization of the corresponding eigenstate. The zeros $\Theta_j$ are
free parameters restricted only by
\be
\Re\, \Theta_j = \frac{\lambda}{2},
\ee
which follows from (iv).

 From the band of next-largest eigenvalues ($\nu = 1$) it is possible to
derive \refto{JohnKM73,KlumSZ89} the correlation length as
\be
\xi = - \frac{2}{\ln k}.  \label{m34}
\ee
 From \refeq{m33} we find numerically $\xi_I = 308.93145$ and
$\xi_{II} = 7105.70704$, for models I and II, respectively, which are
surprisingly large values.

\subsection{Quantum spin chains}

\indent

According to \refeq{m12} there are quantum spin-1 chains associated
with models I and II. After some calculation (see \refto{KlumBiq}) we obtain
\be
H = \sum_{j=1}^N H_{j,j+1},\label{Ham}
\ee
with local interactions:

{Model I}:
\be
H_{j,j+1} = -A_j^2 -2 B^{2}_{j}
- \frac{1}{\sqrt{2}} (A_j B_j + B_j A_j) +
\frac{5}{8}((S^{z}_{j})^2 + (S^{z}_{j+1})^2) + 2 I,
\ee
{Model II}:
\bea
H_{j,j+1} &=&  2B_{j} - \frac{4}{\alpha} A_j^2 - 6 B^{2}_{j}
 - \frac{4}{\sqrt{\alpha}} (A_j B_j + B_j A_j) \non\\
 && +(9 - 4 \alpha)((S^{z}_{j})^2 + (S^{z}_{j+1})^2) - (8 - 8\alpha) I,
\eea
where we have used the definition
\be
A_{j} = S_{j}^{x} S_{j+1}^{x} + S_{j}^{y} S_{j+1}^{y},\quad
B_j = S_{j}^{z} S_{j+1}^{z},\label{abbr}
\ee
in terms of the standard spin operators $S^{x,y,z}$ of spin one, $I$ is the
unit operator, and we have adjusted a scale factor $\tau=8/3$ for model I.

For these Hamiltonians the ground state energy per site
$e_0 = \lim E_{0}/N$ is easily calculated from $\Psi (u)$ (\ref{m21})
\be
\tau e_{0} = -(\ln \Psi)^{\prime} (0),\label{m35}
\ee
and numerically we have $(e_0)_I = -0.338201,\quad (e_0)_{II} = -1.923384$.

More interesting are the low-lying energy-momentum excitations (see
\refto{KlumBiq,KlumZ88App})
\bea
E -E_{0} &=& -{1\over\tau}(\ln l)^{\prime} (0) =
\sum_{j=1}^{\nu} \varepsilon (p_{j}),\non\\
P - P_{0} &=& -i\ln l(0) = \sum_{j=1}^{\nu} p_{j} \label{m36}
\eea
with  energy-momentum dispersion
\be
\varepsilon (p) = \frac{4K}{\tau\pi} \sqrt{(1 - k)^2 + 4k \sin^2 p}.\label{m37}
\ee

Obviously, the Hamiltonian has a gap $\Delta=\varepsilon (0)$ ($\nu=1$)
\be
\Delta = \frac{4K}{\tau\pi} (1 - k) > 0  \label{m38}
\ee
with numerical values $\Delta_{I} = 0.0110033$ and $\Delta_{II} = 0.0018375$
for models I and II, respectively.
The decay of correlations in the groundstate of the Hamiltonian is
described by the correlation length (\ref{m34}).

\section{Solution of models III and IV}

\indent

The crossing symmetry (\ref{m16}) does not hold in the case of
models III and IV. Therefore, the solution in these cases cannot be
obtained by the methods of the previous chapter. However, it is
 straightforward  to see that models III and IV can be mapped
to the six-vertex model, i.e. a two-state vertex model with six
allowed local spin configurations.

The general idea of the mapping is to ignore the signs of all non-zero spin
 values in the allowed configurations of the three-state model.
This is a unique prescription as $b=0$ and $a=c$ for models III and IV,
see (\ref{m3}), (\arabic{modiii}) and (\arabic{modiv}). The remaining
non-zero vertex
configurations and corresponding weights are shown in Fig. 1.
Other configurations than those shown in Fig. 1 are not possible
because $g=0$ for models III and IV. As a spin 1 line could equally well
be a spin $-1$ line, to such a line corresponds an ``internal" degeneracy 2.

\newcounter{cms}
\setlength{\unitlength}{1mm}
\begin{picture}(140, 40)(0,0)
\thicklines
\multiput(12,20)(20, 0){4}{\line(7,0){7}}

\multiput(21,20)(20, 0){4}{\line(7,0){7}}

\multiput(20,12)(20,0){6}{\line(0,7){7}}

\multiput(20,21)(20,0){6}{\line(0,7){7}}

\multiput(18,22)(20,0){4}{\oval(4,4)[br]}

\multiput(22,18)(20,0){4}{\oval(4,4)[tl]}

\multiput(92,20)(20,0){2}{\line(16,0){16}}

\multiput(12,21)(40,0){3}{1}

\multiput(28,21)(60,0){2}{1}

\put(108,21){1}

\multiput(32,21)(40,0){3}{0}

\multiput(48,21)(80,0){2}{0}

\put(68,21){0}

\multiput(22,27)(40,0){2}{1}

\put(122,27){1}

\put(42,27){0}

\multiput(82,27)(20,0){2}{0}

\multiput(22,13)(60,0){2}{1}

\put(122,13){1}

\multiput(42,13)(20,0){2}{0}

\put(102,13){0}
\put(20,2){a}
\put(40,2){d}
\put(60,2){e}
\put(80,2){e}
\put(100,2){p}
\put(120,2){p}
\end{picture}

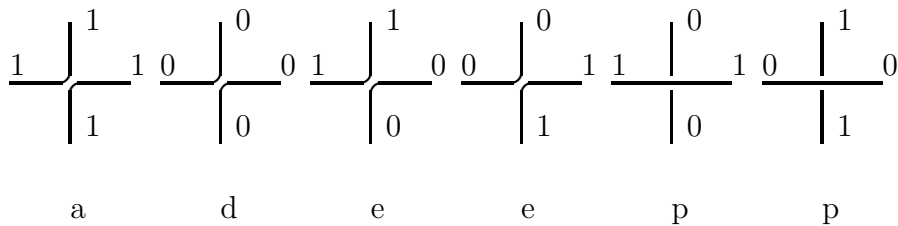
{\small{\bf Fig.\ 1} {The allowed vertices and the Boltzmann
weights of the six-vertex model of interacting spins $\sigma = 0, 1$.
Note that $a$ and $d$ may be different (model IV).}}
\vskip0.5cm

It is clear that the partition functions of the three-state  and the related
two-state model are identical in the thermodynamic limit.
For finite systems the equivalence can be made correct by
introducing a ``seam" of modified weights in the Nth column of the
six-vertex model
\be R_{\nu\, \beta}^{\mu\, \alpha} \rightarrow e^{i\beta \varphi}
 R_{\nu\, \beta}^{\mu\, \alpha}, \label{m41}
\ee
where $\varphi $ is a ``twist" which practically takes any real
value.

The equivalence of the three-state models with periodic boundary
conditions and the six-vertex models with seam are to be understood
 in terms of the spectra of the corresponding transfer matrices
acting on finite chains of length $N$. The action of the transfer
matrices can be thought of as a rearrangement
  of zero-spins  in a background
of non-zero spins without changing the sequence of non-zero spins.
[Note that there are no intersections of lines
of non-zero spins among the allowed vertex configurations.]
Each time a non-zero spin is moved from column $N$ to column 1
the background configuration suffers a shift by one lattice constant
which amounts to multiplying the initial configuration by
$\exp{(i\varphi )}$, where $\varphi$ is the background momentum.
In the reduced description  by the equivalent six-vertex model
there is no such change of the background configuration, but
the factor $\exp{(i\varphi )}$ is imposed by the modified
boundary condition (\ref{m41}). If we denote by $ N_{\pm}, N_{0}$
the numbers of $\pm 1$ and $0$ spins within a row, which are conserved by the
action of the transfer matrices of models III and IV, we have
\bea
N_0 + N_+ + N_- &=& N,\non\\
N_+ - N_- &=& \mbox{magnetization}\; M, \label{m42}\\
\varphi &=& \mbox{multiple of} \frac{2\pi}{N_+ + N_-}.\non
\eea

In the following subsections we treat models III and IV separately
as they correspond to qualitatively different symmetric and
asymmetric six-vertex models, respectively.

\subsection{Model III}

\indent

Model III is defined by the requirement $ a = c = d$, and
  $b = g = 0$, otherwise any values for $a$, $p$  and $e$ are allowed.
The importance of the parameter $\Delta$ \refeq{Del} is known from the
6-vertex model \refto{Baxt82b,LiebWu72}, and the different physical
regions have different parametrizations (\arabic{modiii}).
The associated quantum spin-1 Hamiltonian \refeq{m12} is easily calculated.
It is again of the form \refeq{Ham} with
\be
H_{j,j+1} = - \left[A_j + A_j B_j + B_j A_j\right]
+\Delta \left[(S^z_j)^2 +  (S^z_{j+1})^2 -2B_j^2-1\right],\label{Hamiii}
\ee
in the whole range of $\Delta$ with the abbreviations \refeq{abbr}. In deriving
\refeq{Hamiii} from (\arabic{modiii}) we have adjusted different scale factors
$\tau$ in the three different regions of parametrization.

For $\Delta > 1$ the model possesses simple (frozen in) ferromagnetic order,
the intervals $-1 < \Delta < 1$ and $\Delta < -1$ correspond to the
critical and non-critical antiferromagnetic regimes, respectively.

For $\Delta < -1$, where $\Delta=-\cosh\lambda$ according to \refeq{m8b},
correlation functions decay
with length \refto{JohnKM73,KlumSZ89}
\be
\xi = - \frac{1}{\ln k},  \label{m44}
\ee
where $k$ is the elliptic modulus defined by the requirement
\be
\frac{K^{\prime}}{K} = \frac{2\lambda}{\pi}. \label{m45}
\ee

For $-1 < \Delta < 1$, where $\Delta=-\cos\lambda$,
the 6-vertex model is critical and the
 correlation functions decay
 algebraically. For critical systems the energy levels of
the associated Hamiltonian \refeq{Hamiii} on a
finite chain of length $N$ and periodic boundary conditions
are expected to scale like
\be
E_0=N e_0-{\pi v\over 6N}c,\label{central}
\ee
for the groundstate energy and like
\bea
E-E_0&=&{2\pi\over N}v x,\non\\
P&=&{2\pi\over N}s,\label{finSiz}
\eea
for the excited states where $v$ is the velocity of the
elementary excitations. Systems which are conformally invariant
\refto{BelaPZ84,FrieQS84} have a unique
groundstate and the finite-size amplitudes $c$, $x$, and $s$ are identical to
the central charge of the underlying field theory, the scaling dimensions and
spins of primary fields \refto{Card86a,BlotCN86}.
For instance the asymptotics of two-point
functions are given by
\be
C_r\simeq {1\over r^{2x}}.
\ee

 From the correspondence of model III with a six-vertex model with seam
\refto{KlumBP91,KlumWZ93} we
indeed find \refeq{central} and \refeq{finSiz} to hold with
$c=1$ and
\bea
x  &=& \frac{1 - \lambda/\pi}{2} n^2 + \frac{1}{2(1 - \lambda/{\pi})}
\left(m - \frac{\varphi}
{2\pi} \right)^2+I+{\bar I},\non\\
 s &=& n \left(m - \frac{\varphi}{2\pi}\right)+N{\varphi\over 4\pi}
+I-{\bar I}. \label{m46}
\eea
The groundstate lies in the sector with $N_0=N_++N_-=N/2$ and twist
$\varphi=0$. For the excited states $n=N_0-N/2$, $m$ may take any integer
value and $I$, $\bar I$ are non-negative integers.

However, the numbers $x$ and
$s$ in general do not correspond to scaling dimensions and spins. The problem
can be traced back
to the degeneracy of the groundstate of the system. There are
about $\simeq 2^{N/2}$ background spin configurations for which $N_++N_-=N/2$
with momentum zero. (This implies a residual entropy of ${1\over 2}\ln 2$.)
So, most of the correlation functions are ill-defined if they probe the
background spin configuration, for instance $\langle S_i^z S_j^z\rangle$ or
simply $\langle S^z\rangle$, as any magnetization between $-N/2$ and $+N/2$
may be realized by one of the $\simeq 2^{N/2}$ groundstates. Other correlation
functions are independent of the particular groundstate like
$\langle (S_i^z)^2 (S_j^z)^2\rangle$ and are given by the corresponding
correlation function of the six-vertex model (without twist).

We summarize that only for $\varphi=0$ the set of exponents
of algebraically decaying correlation functions is given by \refeq{finSiz}.
This is a little unfortunate
as for continuous dependence of $x$ on $\varphi$ we could have expected
logarithmic behaviour of the correlations
$C_r\simeq\int d\varphi/r^{2x(\varphi)}\simeq 1/(r^{2x(0)}\ln r)$.

\subsection{Model IV}

\indent

In this case we are dealing with an asymmetric six-vertex model as $a\not= d$
(see Fig.1). The associated quantum spin-1 Hamiltonian is again of the form
\refeq{Ham} and uniformly given by
\be
H_{j,j+1} = - \left[A_j + A_j B_j + B_j A_j\right]
+\alpha_0 \left[1-(S^z_j)^2 -  (S^z_{j+1})^2\right],\label{Hamiv}
\ee
where the coupling $\alpha_0$ takes all real values, $-\infty<\alpha_0
<\infty$. The range $1\le|\alpha_0|$ follows from the parametrization
\refeq{m9a} with $\alpha_0=\pm\coth\lambda$, while the range $|\alpha_0|\le 1$
follows from \refeq{m9b} with $\alpha_0=\pm\tanh\lambda$, and we have adjusted
the scale factors $\tau=\sinh\lambda$ and $\cosh\lambda$ in these two cases,
respectively.

To solve model IV we do not use the parametrization (\arabic{modiv}),
but a different one which is more appropriate to the asymmetric six-vertex
model (and does not satisfy \refeq{m10}). The asymmetric model can
be thought of as a symmetric model in an external magnetic field $B$
acting on horizontal and vertical bond spins, with new weights $\widetilde R$
such that
\be
R^{\mu\;\alpha}_{\nu\;\beta} = \e^{B(\alpha + \beta +
 \mu + \nu)} \widetilde R^{\mu\;\alpha}_{\nu\;\beta}, \label{m47}
\ee
where $(a/d)^{1/4} = \e^B$.

To satisfy \refeq{freeFer} with $e=1$ the $\widetilde R$ can be parametrized
by
\be
\widetilde R^{1\, 1}_{1\, 1} = \widetilde R^{0\, 0}_{0\, 0} = \rho \cos v,\;\;
\widetilde R^{1\, 0}_{0\, 1} = \widetilde R^{0\, 1}_{1\, 0} = \rho,\;\;
\widetilde R^{0\,1}_{0\, 1} = \widetilde R^{1\,0}_{1\,0} = \rho \sin v
\label{m48}
\ee
with fixed scale $\rho = \e^{-2B} $ and an appropriate parameter $v$
which satisfies $0\le{\rm Re }\, v\le{\pi\over 2}$ in the physical region.
The eigenvalues of the transfer matrix of the six-vertex model in external
fields (plus ``twisted boundary conditions") are given by
\be
\Lambda (v) = \rho^N \left(-\e^{2B}\right)^{N-N_0}
\cdot\frac{\e^{2NB+i\varphi} \cos^Nv\,  q(v + \frac{\pi}{2}) + \sin^Nv \,
q(v - \frac{\pi}{2})}{q(v)}, \label{m49}
\ee
where
\be
q(v) = \prod_{j = 1}^{N_0}   \sin (v - v_j) \label{m50}
\ee
is a function of $N_0$ Bethe ansatz numbers $v_j$.
Equation (\ref{m49}) is easily derived
(see for instance \refto{Baxt82b,CPYang67})
as the vertical magnetic field
contributes the overall factor $\e^{2B(N-N_0)}$ and the horizontal field and
 the twist amount to the prefactor $\e^{2BN}\e^{i\varphi}$ of the
 first term on the right hand side of (\ref{m49}).

Inserting \refeq{m50} simplifies \refeq{m49}:
\be
\Lambda (v) =
\left[\e^{2BN+i\varphi} \cos^Nv  + (-1)^{N_0} \sin^Nv \right]
 \, \prod_{j=1}^{N_0}
\left( -\e^{-2B}\frac{\cos (v - v_j)}{\sin (v - v_j)}
\right). \label{m51}
\ee
The factoring is in fact due to the ``free fermion" condition \refeq{freeFer}.
As $\Lambda$ must be analytic at the zeros $v_j$ of $q(v)$, the first factor
must vanish which determines the $v_j$
\be
\tan^N v_j = (-1)^{N_0 + 1} \e^{2BN+i\varphi}, \label{m52}
\ee
which means that the Bethe ansatz equations have decoupled.
Inserting (\ref{m52}) into (\ref{m51}) we obtain the final solution
in terms of the original weights $a$ and $p$
\be
\Lambda = \left[e^{i\varphi} a^N + (-1)^{N_0} p^N \right] \prod_{j=1}^{N_0}
\frac{(1 - p^2)\Theta_j + ap}{a(a - p \Theta_j)} \label{m53},
\ee
where
\be
\Theta_j = \e^{ik_j},\quad k_j = \frac{2\pi}{N} I_j - \frac{\varphi}{N}.
\label{m54}
\ee
The $I_j$ are arbitrary, however distinct integers or half-integers
for odd or even number $N_0$, respectively. Obviously, the influence
of the twist angle $\varphi$ is small, vanishing for $N\to\infty$.

The physical properties are most easily discussed for the associated
Hamiltonian \refeq{Hamiv}. Its eigenvalues follow from \refeq{m12}
\bea
E &=& 2 \sum_{j=1}^{N_0} [\alpha_0 - \cos k_j ] - N \alpha_0,\non\\
P &=& \sum_{j=1}^{N_0} k_j+\varphi,\label{spec}
\eea
for both cases of parametrization (\arabic{modiv}).

The properties depend on the coupling $\alpha_0$.
If $\alpha_0>1$ the groundstate is the $N_0$-vacuum, i.e. the state with
$N_0=0$. With a view to \refeq{Hamiv} this means that all $S_z=1$ or $-1$
along the chain, the degeneracy  is therefore $2^N$.
For $\alpha_0<-1$ the groundstate is the vacuum of non-zero spins. All $S_z$
along the chain are zero, i.e. $N_0=N$ and all $k_j$-modes in \refeq{spec}
are occupied. The state is unique. In both cases we have $E_0=-N |\alpha_0|$
and the gap to the excitations is $\Delta=2(|\alpha_0|-1)$, $N\to\infty$.

For $|\alpha_0|\le 1$ all $k_j$-modes of negative energy in \refeq{spec}
are occupied in the ground state. Its energy is
\be
E_0=-2{\cos{\varphi\over N}\over\sin{\pi\over N}}
\sin{N_0\over N}\pi-(N-2N_0)\alpha_0.
\ee

The model is critical and the underlying conformal
field theory has central charge $c = 1$.
The ``scaling dimensions" and ``spins" (see also subsection 3.1) are given by
\bea
x &=& \frac{{(\Delta n)}^2}{4} + \left(m - \frac{\varphi}{2\pi}\right)^2 + I +
\bar I \non\\
s &=& \Delta n \left(m - \frac{\varphi}{2\pi}\right) +
(N-N_0)\frac{\varphi}{2\pi} + I -
\bar I.
\eea
where $\Delta n$ is the change in the particle number as compared
to the ground state, the integer $m$ describes an asymmetry in the
momentum distribution $k_j$, and $I$, $\bar I$ are non-negative
integers arising from particle-hole excitations.

\section{Conclusion}

\indent

We have studied four exactly solvable three-state vertex models and
corresponding integrable spin-1 chains.

For models I and II the groundstate energy of their associated Hamiltonians,
the elementary excitations and
the correlation lengths could be calculated. These systems turned out to
be non-critical. Unfortunately, the methods of section 2 did not allow for
the calculation of any order parameter or the multiplicity of the groundstate.
However, the patterns of the elementary excitations indicate that model I is an
integrable realization of a Haldane system \refto{Hald83a,Hald83b}
and model II has N\'eel
order. On chains with an even number $N$ of sites the number
$\nu$ of elementary
excitations may be even or odd,
$\nu=$ 1, 2, 3, ... . This excludes dimerization
which would imply $\nu$ even. On chains with an odd number of sites  a similar
analysis to that of section 2 shows that the groundstate of model I is still
separated by a gap from the first band of one-particle excitations, whereas
model II has a groundstate which corresponds to the lowest edge of a
one-particle band. If not conclusive, these findings are at least consistent
with a Haldane system \refto{KlumSZMPG,Frei93} for model I
and N\'eel order for model II.

Models III and IV comprise different physical phases depending on the
value of their coupling parameters $\Delta$ and $\alpha_0$, respectively.
The physical properties were derived by a
mapping to the six-vertex model. We have seen that these three-state models
contain spurious degrees of freedom which essentially do not
participate in the dynamics. This is the reason for the high
degeneracy of the ground states for models I and II (with residual entropy).
It also is the reason why a simple application of conformal invariance
relating finite-size corrections of energy levels to
scaling dimensions fails.

\section*{Acknowledgements}

\indent

One of the authors (S.I.M.) thanks the Institut f\"ur Theoretische
Physik
for its hospitality.

\newpage
\def\and{and\ }
\def\eds{eds.\ }
\def\edi{ed.\ }

\references

\def\mtb{M. T. Batchelor}
\def\rjb{R. J. Baxter}
\def\dk{D. Kim}
\def\pap{P. A. Pearce}
\def\nyr{N. Yu. Reshetikhin}
\def\ak{A. Kl\"umper}

\refis{AbramS64} M. Abramowitz, I. A. Stegun, ``Handbook of
Mathematical Functions", Washington, U.S. National Bureau of Standards 1964;
New York, Dover 1965.

\refis{Affl86} I. Affleck, \prl 56, 746, 1986

\refis{AKLT87} I. Affleck, T. Kennedy, E. H. Lieb \and H. Tasaki, \prl 59,
799, 1987

\refis{AKLT88} I. Affleck, T. Kennedy, E. H. Lieb \and H. Tasaki, \cmp 115,
477, 1988

\refis{AfflGSZ89} I. Affleck, D. Gepner, H. J. Schulz \and T. Ziman,
\jpa 22, 511, 1989

\refis{AkutDW89} Y. Akutsu, T. Deguchi \and M. Wadati, in Braid Group, Knot
Theory and Statistical
Mechanics, \eds C. N. Yang \and M. L. Ge, World Scientific, Singapore, 1989

\refis{AkutKW86a} Y. Akutsu, A. Kuniba \and M. Wadati,\jpj 55, 1466, 1986

\refis{AkutKW86b} Y. Akutsu, A. Kuniba \and M. Wadati,\jpj 55, 2907, 1986

\refis{AlcaBB87} F. C. Alcaraz, M. N. Barber \and \mtb,\prl 58, 771, 1987

\refis{AlcaBB88} F. C. Alcaraz, M. N. Barber \and \mtb,\annp 182, 280, 1988

\refis{AlcaBGR88} F. C. Alcaraz, M. Baake, U. Grimm \and V. Rittenberg,
\jpa 21, L117, 1988

\refis{AlcaM89}  F. C. Alcaraz \and M. J. Martins, \jpa 22, 1829, 1989

\refis{AlcaM90}  F. C. Alcaraz \and M. J. Martins, \jpa 23, 1439-51, 1990

\refis{Alex75} S. Alexander, \pla 54, 353-4, 1975

\refis{AndrBF84} G. E. Andrews, \rjb\ \and P. J. Forrester, \jsp 35, 193,
1984

\refis{And87} P. W. Anderson, Science 235, 1196, 1987

\refis{And90} P. W. Anderson, \prl 64, 1839, 1990

\refis{BDV82} O. Babelon, H. J. de Vega, \and C. M. Viallet, \npb 200 [FS4],
266, 1982

\refis{BarbBP87} \mtb, M.N. Barber \and \pap,\jsp 49, 1117, 1987

\refis{BarbB89} M.N. Barber \and \mtb, \prb 40, 4621, 1989

\refis{Barb91} M.N. Barber, \physica A 170, 221, 1991

\refis{BaresB90} P. A. Bares \and G. Blatter, \prl 64, 2567, 1990

\refis{Bariev8182} R. Z. Bariev, \tmp 49, 261, 1981; 1021, 1982

\refis{Bariev82} R. Z. Bariev, \tmp 49, 1021, 1982

\refis{Bariev91} R. Z. Bariev, \jpa 24, L549, 1991; L919, 1991

\refis{BarievKSZ93} R. Z. Bariev, A. Kl\"{u}mper, A. Schadschneider
\and J. Zittartz, \jpa 26, 1249, 1993; 4863

\refis{Bariev94a} R. Z. Bariev, \prb 49, 1474, 1994

\refis{Bariev94b} R. Z. Bariev, submitted to J. Phys. A

\refis{BarouchM71} E. Barouch \and B. M. McCoy, \pra 3, 786, 1971

\refis{Baxt70} \rjb,\jmp 11, 3116, 1970

\refis{Baxt71b} \rjb,\prl 26, 834, 1971

\refis{Baxt72} \rjb,\annp 70, 193, 1972

\refis{Baxt73} \rjb,\jsp 8, 25, 1973

\refis{Baxt80} \rjb,\jpa 13, L61--70, 1980

\refis{Baxt81a} \rjb,\physica 106A, 18--27, 1981

\refis{Baxt81b} \rjb,\jsp 26, 427--52, 1981

\refis{Baxt82a} \rjb,\jsp 28, 1, 1982

\refis{Baxt82b} \rjb, ``Exactly Solved Models in Statistical Mechanics",
Academic Press, London, 1982.

\refis{BaxtP82} \rjb\space \and \pap,\jpa 15, 897, 1982

\refis{BaxtP83} \rjb\space \and \pap,\jpa 16, 2239, 1983

\refis{BazhR89} V.V. Bazhanov \and \nyr,\ijmpa 4, 115--42, 1989

\refis{BazhB93} V.V. Bazhanov \and \rjb,\physica A 194, 390--396, 1993

\refis{BednorzM86} J. G. Bednorz \and K. A. M"uller, \zpb 64, 189, 1986

\refis{BelaPZ84} A. A. Belavin, A. M. Polyakov \and A. B. Zamolodchikov,
\npb 241, 333, 1984

\refis{Bethe31} H. A. Bethe,\zp 71, 205, 1931

\refis{BlotCN86} H. W. J. Bl\"ote, J. L. Cardy \and M. P. Nightingale, \prl
56, 742,
1986

\refis{BogK89} N. M. Bogoliubov \and V. E. Korepin, \ijmpb 3, 427-439, 1989

\refis{BogIR86} N. M. Bogoliubov, A.\ G.\ Izergin \and
N.\ Y.\ Reshetikhin, \jetpl 44, 405, 1986

\refis{BretzD71} M. Bretz \and J. G. Dash, \prl 27, 647, 1971

\refis{Bretz77} M. Bretz, \prl 38, 501, 1977

\refis{Buy86} W. J. L. Buyers, R. M. Morra, R. L. Armstrong, P. Gerlach
\and K. Hirakawa, \prl 56, 371, 1986

\refis{KawUO89} N. Kawakami, T. Usuki \and A. Okiji, \pla 137, 287, 1989

\refis{KawY90} N. Kawakami \and S.-K. Yang, \prl 65, 2309, 1990

\refis{KawY91} N. Kawakami \and S.-K. Yang, \prb 44, 7844, 1991

\refis{Kaw93} N. Kawakami, \prb 47, 2928, 1993

\refis{KorBI93} V. E. Korepin, N.M. Bogoliubov, \and A.G. Izergin,
``Quantum Inverse Scattering Method and Correlation
Functions", Cambridge University Press, 1993.

\refis{Morra88} R. M. Morra, W. J. L. Buyers, R. L. Armstrong \and K. Hirakawa,
\prb 38, 543, 1988

\refis{ShasS90} B. S. Shastry \and B. Sutherland, \prl 66, 243, 1990

\refis{Shastry88} B. S. Shastry, \jsp 50, 57, 1988

\refis{Stei87} M. Steiner, K. Kakurai, J. K. Kjems, D. Petitgrand \and R. Pynn,
\jappp 61, 3953, 1987

\refis{Tun90} Z. Tun, W. J. L. Buyers, R. L. Armstrong, K. Hirakawa \and
B. Briat, \prb 42, 4677, 1990

\refis{Tun91} Z. Tun, W. J. L. Buyers, A. Harrison \and J. A. Rayne, \prb 43,
13331, 1991

\refis{Ren87} J. P. Renard, M. Verdaguer, L. P. Regnault, W. A. C. Erkelens,
J. Rossa-Mignod \and W. G. Stirling, \eurolett 3, 945, 1987

\refis{Ren88} J. P. Renard, M. Verdaguer, L. P. Regnault, W. A. C. Erkelens,
J. Rossa-Mignod, J. Ribas, W. G. Stirling \and C. Vettier, \jappp 63, 3538,
1988

\refis{Reg89} L. P. Regnault, J. Rossa-Mignod, J. P. Renard, M. Verdaguer
\and C. Vettier, \physica B 156 \& 157, 247, 1989

\refis{Colom87} P. Colombet, S. Lee, G. Ouvrard \and R. Brec, \jcr, 134, 1987

\refis{deGroot82} H. J. M. de Groot, L. J . de Jongh, R. D. Willet \and
J. Reeyk, \jappp 53, 8038, 1982

\refis{Capp88} A. Cappelli, Recent Results in Two-Dimensional Conformal
Field
Theory, in Proceedings of the XXIV International Conference on High Energy
Physics,
\eds R. Kotthaus \and J. K\"uhn, Springer, Berlin, 1988

\refis{CappIZ87a} A. Cappelli, C. Itzykson \and J.-B. Zuber, \npb {280
[FS18]},
445--65, 1987

\refis{CappIZ87b} A. Cappelli, C. Itzykson \and J.-B. Zuber, \cmp 113,
1--26, 1987

\refis{Card84a} J. L. Cardy, \jpa 17, L385, 1984

\refis{Card86a} J. L. Cardy, \npb {270 [FS16]}, 186, 1986

\refis{Card86b} J. L. Cardy, \npb {275 [FS17]}, 200, 1986

\refis{Card88} J. L. Cardy, ``Phase Transitions and Critical Phenomena,
Vol.11",
\eds C. Domb \and J.L. Lebowitz, Academic Press, London 1988

\refis{Card89} J. L. Cardy, Conformal Invariance and Statistical Mechanics,
in Les
Houches, Session XLIV, Fields, Strings and Critical Phenomena, \eds E.
Br\'ezin \and
J. Zinn-Justin, 1989

\refis{ChoiKK90} J.-Y. Choi, K. Kwon \and D. Kim, \eurolett xx, to appear,
1990

\refis{ChoiKP89} J.-Y. Choi, D. Kim \and \pap, \jpa 22, 1661--71, 1989

\refis{CvetDS80} D. M. Cvetkovic, M. Doob \and H. Sachs, ``Spectra of
Graphs", Academic Press, London 1980

\refis{DateJKMO87} E. Date, M. Jimbo, A. Kuniba, T. Miwa, \and M. Okado,
\npb
B290, 231--273, 1987

\refis{DateJKMO88} E. Date, M. Jimbo, A. Kuniba, T. Miwa, \and M. Okado,
\aspm 16,
17, 1988

\refis{DateJMO86} E. Date, M. Jimbo, T. Miwa \and M. Okado,\lmp 12, 209,
1986

\refis{DateJMO87} E. Date, M. Jimbo, T. Miwa \and M. Okado,\prb 35, 2105--7,
1987

\refis{DaviP90} B. Davies \and \pap, \ijmpb {}, this issue, 1990

\refis{deVeK87} H. J. de Vega \and M. Karowski, \npb {285 [FS19]}, 619, 1987

\refis{deVeW85} H. J. de Vega \and F. Woynarovich,\npb 251, 439, 1985

\refis{deVeW90} H. J. de Vega \and F. Woynarovich,\jpa 23, 1613, 1990

\refis{DestdeVeW92} C. Destri \and H. J. de Vega,\prl 69, 2313, 1992

\refis{diFrSZ87} P. di Francesco, H. Saleur \and J.-B. Zuber, \jsp 49,
57--79, 1987

\refis{diFrZ89} P. di Francesco \and J.-B. Zuber, $SU(N)$ Lattice Models
Associated
with Graphs, Saclay preprint SPhT/89-92, 1989

\refis{DijkVV88} R. Dijkgraaf, E. Verlinde \and H. Verlinde, in Proceedings
of the
1987 Copenhagen Conference, World Scientific, 1988

\refis{DijkVVV89} R. Dijkgraaf, C. Vafa, E. Verlinde \and H. Verlinde,\cmp
123, 485, 1989

\refis{DombG76} ``Phase Transitions and Critical Phenomena, Vol.6",
Academic Press, London 1976

\refis{FateZ85} V. A. Fateev \and A. B. Zamolodchikov, \jetp 62, 215, 1985

\refis{FendG89} P. Fendley \and P. Ginsparg, \npb 324, 549--80, 1989

\refis{FodaN89} O. Foda \and B. Nienhuis, \npb {},{},1989

\refis{FoersK93} A. Foerster \and M. Karowski, \npb 408 [FS], 512, 1993

\refis{FrieQS84} D. Friedan, Z. Qiu \and S. Shenker, \prl 52, 1575, 1984; in
``Vertex Operators in Mathematics and Physics", \eds J. Lepowsky, S.
Mandelstam \and
I.M. Singer, Springer, 1984

\refis{FrahmK90} H. Frahm \and V. E. Korepin, \prb 42, 10553, 1990

\refis{FrahmYF90} H. Frahm, N.-C. Yu \and M. Fowler, \npb 336, 396, 1990

\refis{Frei93} W.-D. Freitag, Dissertation, Universit\"at zu K\"oln, 1993

\refis{GepnQ87} D. Gepner \and Z. Qiu, \npb 285, 423--53, 1987

\refis{Gins88} P. Ginsparg,\npb {295 [FS21]}, 153--70, 1988

\refis{Gins89a} P. Ginsparg, Applied Conformal Field Theory, in Les
Houches,
Session XLIV, Fields, Strings and Critical Phenomena, \eds E. Br\'ezin \and
J.
Zinn-Justin, 1989

\refis{Gins89b} P. Ginsparg, Some Statistical Mechanical Models and
Conformal Field
Theories, Trieste Spring School Lectures, HUTP-89/A027

\refis{GradR80} I.S. Gradshteyn \and I.M. Ryzhik, ``Tables of Integrals,
Series and Products", Academic
Press, New York, 1980.

\refis{Grif72} R. B. Griffiths, ``Phase Transitions and Critical Phenomena,
Vol.1",\eds C. Domb \and M. S. Green, Academic Press, London 1972

\refis{Hald83a} F. D. M. Haldane, \prl 50, 1153, 1983

\refis{Hald83b} F. D. M. Haldane, \pla 93, 464, 1983

\refis{Hame85} C. J. Hamer,\jpa 18, L1133, 1985

\refis{Hame86} C. J. Hamer,\jpa 19, 3335, 1986

\refis{Hirsch89a} J. E. Hirsch, \pla 134, 451, 1989

\refis{Hirsch89b} J. E. Hirsch, \physica  C 158, 326, 1989

\refis{Huse82} D. A. Huse,\prl 49, 1121--4, 1982

\refis{Huse84} D. A. Huse, \prb 30, 3908, 1984

\refis{Idz94} M. Idzumi, T. Tokihiro \and M. Arai, \jpI 4, 1151, 1994

\refis{ItzySZ88} C. Itzykson, H. Saleur \and J-B. Zuber, ``Conformal
Invariance and Applications to
Statistical Mechanics", World Scientific, Singapore, 1988

\refis{JimbM84} M. Jimbo \and T. Miwa, \aspm 4, 97--119, 1984

\refis{JimbMO87} M. Jimbo, T. Miwa \and M. Okado, \lmp 14, 123--31, 1987

\refis{JimbMO88} M. Jimbo, T. Miwa \and M. Okado, \cmp 116, 507--25, 1988

\refis{JimbMT89} M. Jimbo, T. Miwa \and A. Tsuchiya,``Integrable Systems in
Quantum Field Theory and
Statistical Mechanics", \aspm 19, ,1989

\refis{JohnKM73} J.D. Johnson, S. Krinsky, \and B.M. McCoy,\pra 8, 2526,
1973

\refis{Kac79} V. G. Kac, \lnp 94, 441--445, 1979

\refis{KadaB79} L. P. Kadanoff \and A. C. Brown, \annp 121, 318--42, 1979

\refis{Karo88} M. Karowski, \npb {300 [FS22]}, 473, 1988

\refis{Kato87} A. Kato, \mpla 2, 585, 1987

\refis{KimP87} \dk\space \and \pap,\jpa 20, L451--6, 1987

\refis{KimP89}  \dk\space \and \pap,\jpa 22, 1439--50, 1989

\refis{Kiri89} E. B. Kiritsis, \plb  217, 427, 1989

\refis{KiriR86} A. N. Kirillov \and N. Yu. Reshetikhin,\jpa 19, 565, 1986

\refis{KiriR87} A. N. Kirillov \and N. Yu. Reshetikhin,\jpa 20, 1565, 1987

\refis{KlassM90} T. R. Klassen \and E. Melzer, \npb 338, 485, 1990

\refis{KlassM91} T. R. Klassen \and E. Melzer, \npb 350, 635, 1991

\refis{KlumBiq} A. Kl\"{u}mper, \eurolett 9, 815, 1989; \jpa 23, 809, 1990

\refis{KlumB90} A. Kl\"{u}mper \and \mtb,\jpa 23, L189, 1990

\refis{KlumBP91} A. Kl\"{u}mper, \mtb \ \and \pap, \jpa 24, 3111--3133, 1991

\refis{KlumP91} A. Kl\"{u}mper \and \pap, \jsp 64, 13--76, 1991

\refis{Klum92c} A. Kl\"{u}mper, unver"offentlichte Rechnungen, (1992)

\refis{KlumZ88} A. Kl\"{u}mper \and J. Zittartz,\zpb 71, 495, 1988

\refis{KlumZ88App} A. Kl\"{u}mper \and J. Zittartz,\zpb 71, 495, 1988,
Appendix A

\refis{KlumZ89} A. Kl\"{u}mper \and J. Zittartz,\zpb 75, 371, 1989

\refis{KlumZ8VM} A. Kl\"{u}mper \and J. Zittartz,\zpb 71, 495, 1988;
\zpb 75, 371, 1989

\refis{KlumSZ89} A. Kl\"{u}mper, A. Schadschneider \and J. Zittartz,
\zpb 76, 247, 1989

\refis{KlumSZMPG} A. Kl\"{u}mper, A. Schadschneider \and J. Zittartz,
\jpa 24, L955-L959, 1991; \zpb 87, 281-287, 1992

\refis{Klum89} \ak, \eurolett 9, 815, 1989

\refis{KlumP92} \ak\  \and \pap, \physica 183A, 304-350, 1992

\refis{Klum92} \ak , \Annp 1, 540, 1992

\refis{Klum93} \ak , \zpb 91, 507, 1993

\refis{Klum92b} \ak, in preparation

\refis{KlumWZ93} \ak, T. Wehner \and J. Zittartz, \jpa 26, 2815, 1993

\refis{Klum94} \ak , in Vorbereitung

\refis{KlumWeh94} \ak\ \and T. Wehner, in Vorbereitung

\refis{Knabe88} S. Knabe, \jsp 52, 627, 1988

\refis{Koma} T. Koma, \ptp 78, 1213, 1987; \bf 81, \rm 783, (1989)

\refis{KorepinS90} V. E. Korepin \and N. A. Slavnov, \npb 340, 759, 1990

\refis{ItsIK92} A. R. Its, A. G. Izergin \and V. E. Korepin, \physica D 54,
351, 1992

\refis{ItsIKS93} A. R. Its, A. G. Izergin, V. E. Korepin \and N. A. Slavnov,
\prl 70, 1704, 1993

\refis{IKR89} A. G. Izergin, V. E. Korepin \and N. Yu. Reshetikhin, \jpa 22,
2615, 1989

\refis{KuniY88} A. Kuniba \and T. Yajima,\jsp 52, 829, 1988

\refis{KuliRS81} P. P. Kulish, N. Yu. Reshetikhin \and E. K. Sklyanin, \lmp
5, 393, 1981

\refis{Kuniba92} A. Kuniba, ``Thermodynamics of the $U_q(X_r^{(1)})$ Bethe
Ansatz System with $q$ a Root of Unity", ANU preprint (1991)

\refis{LeeS88} K. Lee \and P. Schlottmann, \jpcoll 49 C8, 709, 1988

\refis{Lewi58} L. Lewin, Dilogarithms and Associated Functions, MacDonald,
London, 1958

\refis{LiebWu68} E. H. Lieb \and F. Y. Wu, \prl 20, 1445, 1968

\refis{LiebWu72} E. H. Lieb \and F. Y. Wu,
``Phase Transitions and Critical Phenomena,
Vol.1",
\eds C. Domb \and M. S. Green, Academic Press, London 1988

\refis{LutherP74} A. Luther \and I. Peschel, \prb 9, 2911, 1974

\refis{Martins91} M. J. Martins, \prl 22, 419, 1991 and private communication
(1991)

\refis{Muell} E. M\"uller-Hartmann, unpublished results, (1989)

\refis{Muell89} E. M\"uller-Hartmann, unver"offentlichte Ergebnisse, (1989)

\refis{Mura89} J. Murakami, \aspm 19, 399--415, 1989

\refis{Nien87} B. Nienhuis, in Phase Transitions and Critical Phenomena,
Vol.11,
\eds C. Domb \and J.L. Lebowitz, Academic Press, 1987

\refis{NahmRT92} W. Nahm, A. Recknagel \and M. Terhoven, Preprint ``Dilogarithm
identities in conformal field theory'', 1992

\refis{NighB86} M. P. Nightingale \and H. W. J. Bl"ote, \prb 33, 659, 1986

\refis{OwczB87} A. L. Owczarek \and \rjb,\jsp 49, 1093, 1987

\refis{ParkBiq} J. B. Parkinson,\jpc 20, L1029, 1987; \jpc 21, 3793, 1988;
\jphc 8, 1413, 1988

\refis{ParkB85} J. B. Parkinson \and J. C. Bonner, \prb 32, 4703, 1985

\refis{PaczP90} I. D. Paczek \and J. B. Parkinson,\jpcon 2, 5373, 1990

\refis{Pasq87a} V. Pasquier,\npb {285 [FS19]}, 162, 1987

\refis{Pasq87b} V. Pasquier,\jpa 20, {L217, L221}, 1987

\refis{Pasq87c} V. Pasquier,\jpa 20, {L1229, 5707}, 1987

\refis{Pasq88} V. Pasquier,\npb {B295 [FS21]}, 491--510, 1988

\refis{Pear85} \pap,\jpa 18, 3217--26, 1985

\refis{Pear87prl} \pap,\prl 58, 1502--4, 1987

\refis{Pear87jpa} \pap,\jpa 20, 6463--9, 1987

\refis{Pear90ijmpb} \pap,\ijmpb 4, 715--34, 1990

\refis{PearB90} \pap\space \and \mtb, \jsp 60, 77--135, 1990

\refis{PearK87} \pap \and \dk, \jpa, 20, 6471-85, 1987

\refis{PearS88} \pap\space \and K. A. Seaton,\prl 60, 1347, 1988

\refis{PearS89} \pap\space \and K. A. Seaton,\annp 193, 326, 1989

\refis{PearS90} \pap\space \and K. A. Seaton,\jpa 23, 1191--1206, 1990

\refis{Pear91} \pap, Row Transfer Matrix Functional Equations for
$A$--$D$--$E$ Lattice Models,
 to be published, 1991

\refis{PearK91} \pap\space \and A. Kl\"umper, \prl 66, 974, 1991

\refis{Pear92} \pap, \ijmpa 7, Suppl.1B, 791, 1992

\refis{PerkS81} J. H. H. Perk \and C. L. Schultz, \pla 84, 407, 1981

\refis{Resh83jetp} \nyr,\jetp 57, 691, 1983

\refis{Resh83lmp} \nyr, \lmp 7, 205--13, 1983

\refis{Sale88} H. Saleur, Lattice Models and Conformal Field Theories, in
Carg\`ese
School on Common Trends in Condensed Matter and Particle Physics, 1988

\refis{SaleB89} H. Saleur \and M. Bauer, \npb 320, 591--624, 1989

\refis{SaleD87} H. Saleur \and P. di Francesco, Two Dimensional Critical
Models on a
Torus, in Brasov Summer School on Conformal Invariance and String Theory,
1987

\refis{Samuel73} E. J. Samuelson, \prl 31, 936, 1973

\refis{Schlott87} P. Schlottmann, \prb 36, 5177, 1987

\refis{Schlott92} P. Schlottmann, \jpc 4, 7565, 1992

\refis{Schul83} C. L. Schultz, \physica 122A, 71, 1983

\refis{SeatP89} K. A. Seaton \and \pap\space, \jpa 22, 2567--76, 1989

\refis{Strog79} Yu. G. Stroganov, \pla 74, 116, 1979

\refis{Suth70} B. Sutherland, \jmp 11, 3183, 1970

\refis{Suth75} B. Sutherland, \prb 12, 3795, 1975

\refis{Suzuki85} M. Suzuki, \prb 31, 2957, 1985

\refis{SuzukiI87} M. Suzuki \and M. Inoue, \ptp 78, 787, 1987

\refis{Suzuki87} M. Suzuki, in ``Quantum Monte Carlo Methods in
Equilibrium and Nonequilibrium Systems",
\edi M. Suzuki, Springer Verlag, 1987

\refis{SuzukiAW90} J. Suzuki, Y. Akutsu \and M. Wadati, \jpj 59, 2667-2680,
1990

\refis{SuzukiNW92} J. Suzuki, T. Nagao \and M. Wadati, \ijmpb 6, 1119, 1992

\refis{Tak71} M. Takahashi, \ptp 46, 401, 1971

\refis{TakTBA} M. Takahashi, \ptp 46, 401, 1971; \ptp 50, 1519, 1973

\refis{Tak91} M. Takahashi, \prb 43, 5788, 1991; \prb 44, 12382, 1991

\refis{Tak91a} M. Takahashi, \prb 43, 5788, 1991

\refis{Tak91b} M. Takahashi, \prb 44, 12382, 1991

\refis{TempL71} H. N. V. Temperley \and E. H. Lieb, \prs 322, 251, 1971

\refis{Tetel82} M. G. Tetel'man, \jetp 55, 306, 1982

\refis{TruS83} T. T. Truong \and K. D. Schotte, \npb 220, 77, 1983

\refis{Tsun91} H. Tsunetsugu, \jpj 60, 1460, 1991

\refis{vonGR87} G. von Gehlen \and V. Rittenberg, \jpa 20, 227, 1987

\refis{WadaDA89} M. Wadati, T. Deguchi \and Y. Akutsu, \prep 180, 247--332,
1989

\refis{Woyn87} F. Woynarovich, \prl 59, 259, 1987

\refis{WoynE87} F. Woynarovich \and H.-P. Eckle, \jpa 20, L97, 1987

\refis{Yang69} C. N. Yang \and C. P. Yang, \jmp 10, 1115, 1969

\refis{Yang66} C. N. Yang \and C. P. Yang, \pr 147, 303, 1966; 150, 321

\refis{Yang62} C. N. Yang, \rmp 34, 691, 1962

\refis{Yang67} C. N. Yang, \prl 19, 1312, 1967

\refis{CPYang67} C. P. Yang, \prl 19, 586, 1967

\refis{YangG89} C. N. Yang \and M. L. Ge (Editors), Braid Group, Knot Theory
and Statistical
Mechanics, World Scientific, Singapore, 1989

\refis{ZamoF80} A. B. Zamolodchikov \and V. Fateev, \sjnp 32, 198, 1980

\refis{Zamo80} A. B. Zamolodchikov, \jetp 52, 325, 1980; \cmp 79, 489, 1981

\refis{Zamo91} Al. B. Zamolodchikov, \plb 253, 391--4, 1991; \npb 358,
497--523, 1991

\refis{Zamo91a} Al. B. Zamolodchikov, \plb 253, 391--4, 1991

\refis{Zamo91b} Al. B. Zamolodchikov, \npb 358, 497--523, 1991

\refis{ZhangR88} F. C. Zhang and T. M. Rice, \prb 37, 3759, 1988

\endreferences

\end{document}